\documentclass[conference]{IEEEtran}
\usepackage{graphicx}
\usepackage{amssymb}
\usepackage{bm}
\usepackage{amsmath}
\usepackage{pgfplots}
\usepackage{balance}
\usepackage{epstopdf}
\usepackage{algpseudocode}
\usepackage{algorithm,algpseudocode}
\usepackage{algorithmicx}
\usepackage[noadjust]{cite}
\usepackage{booktabs} 
\pgfplotsset{compat=newest}
\pgfplotsset{plot coordinates/math parser=false}
\newlength\figureheight
\newlength\figurewidth

\usepackage{color,soul}
\IEEEoverridecommandlockouts

\ifCLASSOPTIONcompsoc
    \usepackage[caption=false, font=normalsize, labelfont=sf, textfont=sf]{subfig}
\else
\usepackage[caption=false, font=footnotesize]{subfig}
\fi
  
\usepackage{amsmath,amssymb}
\usepackage{mathtools} 
\usepackage{xspace}

\newcommand{\Eqref}[1]{(\ref{#1})}

\newcommand{\Figref}[1]{Fig.~\ref{#1}}

\newcommand{\Thmref}[1]{Theorem~\ref{#1}}

\newcommand{\herm}{^\text{H}}

\newcommand{\trans}{^\text{T}}
\newcommand{\inv}{^{-1}}

\newcommand{\bx}{\mathbf{x}}
\newcommand{\bz}{\mathbf{z}}

\newcommand{\bs}{\mathbf{s}}

\newcommand{\bH}{\mathbf{H}}
\newcommand{\bh}{\mathbf{h}}

\newcommand{\bg}{\mathbf{g}}

\newcommand{\bw}{\mathbf{w}}
\newcommand{\bY}{\mathbf{Y}}

\newcommand{\bI}{\mathbf{I}}

\newcommand{\be}{\mathbf{e}}

\newcommand{\bU}{\mathbf{U}}
\newcommand{\bu}{\mathbf{u}}

\newcommand{\bN}{\mathbf{N}}

\newcommand{\bPhi}{\mathbf{\Phi}}
\newcommand{\bphi}{\boldsymbol{\phi}}

\newcommand{\bzero}{\boldsymbol{0}}
\newcommand{\mC}{\mathbb{C}}

\newcommand{\CN}{\mathcal{CN}}
\newcommand{\tc}{\tau_{\textsc{c}}}

\newcommand{\boundellipse}[3]
{(#1) ellipse [x radius=#2,y radius=#3]
}

\newtheorem{theorem}{Theorem}

\newtheorem{corollary}{Corollary}

\newtheorem{remark}{Remark}

\DeclareMathOperator{\E}{\mathsf{E}}

\newcommand{\sigmash}{\sigma_{\textrm{sh}}}

\newcommand{\EX}[1]{\mathsf{E}\left\{{#1}\right\}}

\newcommand{\CG}[2]{\mathcal{CN}\left({#1},{#2}\right)}

\newcommand{\B}[1]{{\mathbf{#1}}}

\newcommand{\cP}{\mathcal{P}}

\newcommand{\norm}[1]{{ \left\Vert #1 \right\Vert }}

\newcommand{\ik}{i_{k}}

\newcommand{\tp}{\tau_{\textsc{p}}}
\newcommand{\td}{\tau_{\textsc{d}}}

\makeatletter
\def\@setsize#1#2#3#4{
    \@nomath#1
    \let\@currsize#1
    \baselineskip #2
    \baselineskip \baselinestretch\baselineskip
    \parskip \baselinestretch\parskip
    \setbox\strutbox \hbox{
        \vrule height.7\baselineskip
            depth.3\baselineskip
            width\z@}
    \skip\footins \baselinestretch\skip\footins
    \normalbaselineskip\baselineskip#3#4}
\makeatother

\makeatletter
\newcommand{\setstretch}[1]{
    \def\baselinestretch{#1}%
    \@currsize
    }
\makeatother



\begin{document}

\begin{figure*}[t!]
\normalsize
Paper published in 2018 IEEE Global Conference on Signal and Information Processing (GlobalSIP). \\Added to IEEE Xplore: February 21, 2019. DOI: 10.1109/GlobalSIP.2018.8646666. 

\

\textcopyright~2019 IEEE. Personal use of this material is permitted.  Permission from IEEE must be obtained for all other uses, in any current or future media, including reprinting/republishing this material for advertising or promotional purposes, creating new collective works, for resale or redistribution to servers or lists, or reuse of any copyrighted component of this work in other works.
\vspace{17cm}
\end{figure*}

\newpage

\title{Downlink Spectral Efficiency of Cell-Free Massive MIMO with Full-Pilot Zero-Forcing}
\author{\IEEEauthorblockN{Giovanni Interdonato$^{*\dagger}$, Marcus Karlsson$^\dagger$, Emil Bj\"{o}rnson$^\dagger$, Erik G. Larsson$^\dagger$}
\IEEEauthorblockA{$^*$Ericsson Research, 581 12 Link\"oping, Sweden\\
$^\dagger$Department of Electrical Engineering (ISY), Link\"oping University, 581 83 Link\"oping, Sweden\\
giovanni.interdonato@ericsson.com, \{marcus.karlsson, emil.bjornson, erik.g.larsson\}@liu.se
\thanks{This paper was supported by the European Union's Horizon 2020 research
and innovation programme under grant agreement No 641985 (5Gwireless).}}}
\maketitle

\begin{abstract}
Cell-free Massive multiple-input multiple-output (MIMO) ensures ubiquitous communication at high spectral efficiency (SE) thanks to increased macro-diversity as compared cellular communications. However, system scalability and performance are limited by fronthauling traffic and interference. Unlike conventional precoding schemes that only suppress intra-cell interference, \textit{full-pilot zero-forcing} (fpZF), introduced in~\cite{Bjornson2016a}, actively suppresses also inter-cell interference, without sharing channel state information (CSI) among the access points (APs). In this study, we derive a new closed-form expression for the downlink (DL) SE of a cell-free Massive MIMO system with multi-antenna APs and fpZF precoding, under imperfect CSI and pilot contamination. The analysis also includes max-min fairness DL power optimization. Numerical results show that fpZF significantly outperforms maximum ratio transmission scheme, without increasing the fronthauling overhead, as long as the system is sufficiently distributed. 
\end{abstract}

\begin{IEEEkeywords}
Cell-free Massive MIMO, full-pilot zero-forcing, downlink spectral efficiency, max-min fairness power control. 
\end{IEEEkeywords}

\section{Introduction}

Cell-free Massive MIMO~\cite{Ngo2017b} refers to a time-division duplex (TDD) distributed Massive MIMO system where all the APs coherently serve all the users (UEs) in the same time-frequency resources. Each AP is connected to a central processing unit (CPU), through a fronthaul network, that is responsible for the coordination. Cell-free Massive MIMO can be seen as the scalable implementation of coordinated multi-point with joint transmission (CoMP-JT)~\cite{Boldi2011a} as channel estimation and precoding is performed locally at each AP by leveraging the \textit{channel reciprocity} of such a TDD system. Moreover, cell-free Massive MIMO is implemented in a user-centric fashion \cite{Bjornson2013d}: a given UE is served by all the APs it is able to reach. As a result, each UE experiences no cell boundaries in the data transmission, hence the terminology cell-free. The very large total number of antennas in cell-free Massive MIMO leads to an increased macro-diversity and \textit{favorable propagation}, namely the UEs are easier to be spatially separated resulting in negligible inter-user interference.


Cell-free Massive MIMO, in its canonical form~\cite{Ngo2017b}, consists in single-antenna APs and maximum ratio transmission (MRT) precoding. 
Recent works~\cite{Nayebi2017a,Ngo2018a} extended the analysis to multi-antenna APs and zero-forcing (ZF) precoding. However, implementing ZF requires instantaneous CSI to be sent from the APs to the CPU, where the ZF precoder is calculated and fed back. This might result in unmanageable fronthauling traffic, performance degradation and unscalable architecture when the number of antennas and UEs grows. Conversely, fpZF scheme, introduced in~\cite{Bjornson2016a}, suppresses interference in a fully distributed, coordinated and scalable fashion. In~\cite{Chien2016a}, fpZF performance is evaluated in multi-cell co-located massive MIMO (i.e., non-coherent transmission).   

\textbf{Contributions:} In this study, we evaluate the performance of a cell-free Massive MIMO system with multi-antenna APs and fpZF precoding. We derive a new closed-form expression for the DL SE, under i.i.d. Rayleigh fading, addressing imperfect CSI and pilot contamination. Our analysis also includes max-min fairness DL power optimization under per-AP power constraints. A performance comparison is carried out against cell-free Massive MIMO implementing MRT, both for single and multi-antenna APs. This comparison is restricted to precoding schemes not requiring CSI sharing, and the performance are evaluated at the corresponding optimal operation point of all the schemes, according to the max-min fairness policy.        


\section{System Model}
\label{sec:system-model}

We analyze a cell-free Massive MIMO system operating in TDD in which the single-antenna UEs are jointly served by all the APs. Let $L$, $M$, $K$ be the number of APs, antenna elements per AP, and active UEs, respectively, with $LM \gg K$. 

We consider a standard block-fading channel model where  $\B{h}_{l,k}\!\in\!\mC^{M\times 1}$ is the channel response between the $k$th UE and the $l$th AP, $k=1,\ldots,K$, $l=1,\ldots,L$. The channel responses are independent identically distributed (i.i.d.) in each coherence interval, $\B{h}_{l,k}\!\sim\!\mathcal{CN}(\B{0},\beta_{l,k}\B{I}_M)$, where $\beta_{l,k}$ is the large-scale fading coefficient. We assume that each AP knows the large-scale fading coefficients associated with itself.


The coherence interval is $\tc$ symbols long and we use $\tp$ of these on \textit{pilots}, leaving $\td=\tc-\tp$ for payload data. Let $\xi^{\mathrm{DL}}$ be the fraction of the data symbols that are used for DL payload transmission, hence $0 < \xi^{\mathrm{DL}} \leq 1$. The payload part will be split between DL and uplink (UL) transmission with length $\xi^{\mathrm{DL}}\td$ and $(1-\xi^{\mathrm{DL}})\td$ symbols, respectively.

\subsection{Uplink Training and Channel Estimation}
\label{subsec:ULtraining}

In a pilot-based UL training, all the UEs synchronously send their pilot sequences 
to the APs, once per coherence interval. These pilots enable the APs to estimate the channels.

%

We assume that $\tp$ orthogonal pilots are available, $\tp \leq K$. Let $i_k\in \{1,\dots,\tp\}$ be the index of the pilot used by UE $k$, $\bphi_{i_k} \in \mC^{\tp\times1}$ is the pilot sequence sent by the $k$th UE. We define $\cP_k \subset\{1,\ldots,K\}$ as the set of indices, including $k$, of UEs that transmit the same pilot as UE $k$, hence $\ik~=~i_t~\Leftrightarrow~t~\in \cP_{k}$. The pilot sequences are mutually orthogonal and normalized such that
\begin{align}
\bphi_{i_t}\herm \bphi_{i_k} = \begin{cases}
	0, & t \notin \cP_k \\
	\tp, & t \in \cP_k.
	\end{cases}
\end{align}
The pilot signal received at AP $l$ 
is given by
\begin{equation}\label{eq:uplink:received:pilots}
 \bY_l \triangleq \sum\limits_{k=1}^{K}\bh_{l,k}\sqrt{p_k}\bphi_{i_k}\herm + \bN_{l} \in \mC^{M\times\tp},
\end{equation}
where $\bN_{l} \in \mC^{M\times\tp}$ is a Gaussian noise matrix with i.i.d.~$\CG{0}{1}$ elements, and $p_k$ is the UL normalized transmit power. The energy spent on pilot by UE $k$ is $p_k\tp$. 

The minimum mean square error (MMSE) estimate of the channel between UE $k$ and AP $l$ is, according to~\cite{Chien2016a}, 
\begin{equation} \label{eq:channel-estimate}
\hat{\bh}_{l,k} \triangleq c_{l,k}\bY_{l} \bphi_{\ik},
\end{equation}
where we have defined
\begin{equation}
c_{l,k} \triangleq  \dfrac{\sqrt{p_k}\beta_{l,k}}{\tp \sum_{t\in\cP_k}p_{t}\beta_{l,t} + 1}.
\end{equation}  
The channel estimate and estimation error, denoted by $\hat{\bh}_{l,k}$ and $\tilde{\bh}_{l,k}$, respectively, with ${\bh}_{l,k} = \hat{\bh}_{l,k} + \tilde{\bh}_{l,k}$, are distributed as $\hat{\bh}_{l,k}\sim\CN(\bzero, \gamma_{l,k}\bI_{M})$, $\tilde{\bh}_{l,k}\sim\CN(\bzero, (\beta_{l,k}-\gamma_{l,k})\bI_{M})$, where
\begin{equation}
\gamma_{l,k} \triangleq \dfrac{p_k\tp\beta_{l,k}^{2}}{\tp\sum_{t\in\cP_k}p_t\beta_{l,t} + 1}.
\end{equation} 
\begin{remark}[Pilot contamination]
The channel estimates to two different UEs using the same pilot sequence are parallel. For any pair of UEs $k$ and $t$, with $t \in \cP_k$, $t \neq k$, the respective channel estimates to any AP $l$ are linearly dependent as  
\begin{equation} \label{eq:parallel-estimates}
\hat{\bh}_{l,k} = \dfrac{\sqrt{p_k}\beta_{l,k}}{\sqrt{p_t}\beta_{l,t}}\hat{\bh}_{l,t}.
\end{equation}
Hence, the AP cannot spatially separate the UEs sharing the same pilot and cannot suppress the corresponding interference.
\end{remark} 

\subsection{Downlink Data Transmission}

The APs use the channel estimates to perform fpZF precoding. Unlike canonical ZF, fpZF does not require  CSI transmission from all the APs to the CPU, as each AP constructs its precoders by using only its local CSI. Hence, fpZF has the same fronthaul requirements as MRT. 


If $\tp < K$, some of the estimated channels are parallel, thus $\hat{\bH}_l = [\hat{ \bh }_{l,1}, \ldots, \hat{\bh}_{l,K}]~\in~\mC^{ M \times K}$ is rank-deficient. To define the fpZF precoder at AP $l$, we construct the full-rank matrix
\begin{equation}
\bar{\bH}_l \triangleq \bY_{l}\bPhi \in \mC^{M\times \tp},
\end{equation}
which is connected to the respective channel estimate by
\begin{equation} \label{eq:hat-to-bar-h}
\hat{\bh}_{l,k} = c_{l,k}\bar{\bH}_l\be_{\ik},
\end{equation} 
where $\bPhi\!=\![\bphi_1,\ldots,\bphi_{\tp}]\!\in\!\mC^{\tp\!\times\!\tp}$, and $\be_{\ik}$ denotes the $\ik$th column of $\bI_{\tp}$.

The precoding vector used by AP $l$ and intended for UE $k$, $ \mathbf{w}_{l,\ik} \in \mC^{M \times 1}$,  can be written as
\begin{equation}  \label{eq:Linear-Precoding-Vector}
\bw_{l,\ik}  = \frac{ \bar{\bH}_{l} \left(\bar{\bH}_l\herm\bar{\bH}_l\right)\inv \be_{\ik} }
{\sqrt{ \EX{ \norm{\bar{\bH}_{l} \left(\bar{\bH}_l\herm\bar{\bH}_l\right)\inv \be_{\ik}}^{2}} }}.
\end{equation}
\begin{remark}
Each AP has $\tp$ precoding vectors, one per pilot. The same vector is used for all the UEs sharing the same pilot.
\end{remark}

The data signal $\mathbf{x}_l$, transmitted by AP $l$, is  
\begin{equation}
\bx_l = \sum_{k=1}^{K} \sqrt{\rho_{l,k}} \mathbf{w}_{l,\ik} q_k,
\end{equation}
where $\rho_{l,k}$ is the transmit power allocated to UE $k$, and $q_k$ is the independent data symbol intended for UE $k$, having unit power, $\E\{ | q_{k}|^2 \} = 1$. 
The received signal at the UE $k$ is 
\begin{equation} \label{eq:Downlink-Signal}
y_k\!=\!\sum_{l = 1}^{L}\!\sqrt{ \rho_{l,k} } \bh_{l,k}\herm \bw_{l,\ik} q_k \!+\!\sum\limits_{l=1}^{L}\!\sum_{ t \neq k }^{K}\!\sqrt{ \rho_{l,t} } \bh_{l,k}\herm \bw_{l,i_t} q_t\!+\!n_k. 
\end{equation}
The first term in \eqref{eq:Downlink-Signal} is the desired signal for the $k$th UE, the second term represents the inter-user interference, and the third term is the independent noise $n_k\sim\CG{0}{1}$ at UE $k$.

\section{Performance Analysis} 
\subsection{Achievable Downlink Spectral Efficiency} \label{subsec:DLrate}
The expression in~\eqref{eq:Downlink-Signal} can be rewritten as
\begin{align}
\label{eq:signalUE}
y_k = \text{CB}_k \cdot q_k + \text{BU}_k \cdot q_k + \sum_{ t \neq k }^{K} \text{UI}_{kt} \cdot q_t + n_k,  
\end{align}  
where $\text{CB}_k$, $\text{BU}_k$, and $\text{UI}_{kt}$ reflect the coherent beamforming gain, beamforming gain uncertainty, 
and inter-user interference, respectively, given by 
\begin{align}
& \text{CB}_k \triangleq \sum_{l=1}^L \sqrt{\rho_{l,k}} \EX{\bh_{l,k}\herm \bw_{l,\ik}}, \label{eq:DS}\\
& \text{BU}_k \triangleq \sum_{l=1}^L \left( \sqrt{\rho_{l,k}} \bh_{l,k}\herm \bw_{l,\ik}-\sqrt{\rho_{l,k}} \EX{\bh_{l,k}\herm \bw_{l,\ik}}\right), \label{eq:BU}\\
& \text{UI}_{kt} \triangleq \sum\limits_{l=1}^{L} \sqrt{\rho_{l,t}} \bh_{l,k}\herm \bw_{l,i_t}.
\label{eq:UI}
\end{align}  


UE $k$ in \Eqref{eq:signalUE} effectively sees a deterministic channel ($\text{CB}_k$) with some uncorrelated noise. 
By invoking the arguments from \cite{Medard2000a} or \cite[Sec. 2.3.2]{Marzetta2016a}, an achievable DL SE, for UE $k$, can be written as stated in \Thmref{Theorem-Lower-Bound-Rate}.

\begin{theorem} \label{Theorem-Lower-Bound-Rate}
A lower bound on the DL ergodic capacity of an arbitrary UE $k$ is given by 
\begin{equation} \label{eq:Sum-Rate-k}
\mathrm{SE}_k = \xi^{\mathrm{DL}} \left( 1 - \frac{\tp}{\tc} \right) \log_2 (1 + \mathrm{SINR}_k ) \quad \textrm{[bit/s/Hz]},
\end{equation}
where $\mathrm{SINR}_k$ (signal-to-interference-plus-noise ratio) is 
\begin{equation}
\label{eq:sinr:dl}
\frac{\!\left|\sum\limits_{l=1}^L\!\sqrt{\rho_{l,k}}\EX{\bh_{l,k}\herm \bw_{l,\ik}\!}\!\right|^2\!}{\!\sum\limits_{t=1}^K\!\EX{\!\left| \sum\limits_{l=1}^L\!\sqrt{\rho_{l,t}}\bh_{l,k}\herm\!\bw_{l,i_t}\!\right|^2\!}\!-\!\left|\!\sum\limits_{l=1}^L\!\sqrt{\rho_{l,k}}\EX{\bh_{l,k}\herm\!\bw_{l,\ik}\!}\!\right|^2\!+\!1}.
\end{equation}
\end{theorem}

\subsection{Achievable Downlink Spectral Efficiency for i.i.d. Rayleigh Fading and Full-Pilot Zero-Forcing Precoding}
To calculate the effective $\mathrm{SINR}_k$ in Theorem~\ref{Theorem-Lower-Bound-Rate}, we first find a simple expression for the inner product $\hat{\bh}_{l,k}\herm \bw_{l,i_t}$. The normalization term in~\Eqref{eq:Linear-Precoding-Vector} is given by \cite[Lemma 2.10]{Tulino2004a}
\begin{equation} \label{eq:den-ZF-definition}
 \EX{ \norm{\bar{\bH}_{l} \left(\bar{\bH}_l\herm\bar{\bH}_l\right)\inv \be_{\ik}}^{2}} = \dfrac{c^2_{l,k}}{\gamma_{l,k}(M-\tp)}.
\end{equation}
fpZF precoding has the ability to suppress interference towards all the UEs unless they share the same pilot sequence: 
\begin{align} \label{eq:ZF-Property}
\hat{\bh}_{l,k}\herm \bw_{l,i_t} &= \dfrac{c_{l,k}}{c_{l,t}}\be_{i_k}\herm\be_{i_t}\sqrt{\gamma_{l,t}(M-\tp)} \nonumber \\
&=
\begin{cases}  0, &  t \notin \cP_k, \\
\sqrt{\gamma_{l,k}(M-\tp)}, & t \in \cP_k. \end{cases}
\end{align}
By substituting~\eqref{eq:ZF-Property} into~\eqref{eq:sinr:dl}, and computing the expected values, the ergodic SE is obtained in closed form. 
\begin{corollary} \label{ZF-DL-Rate}
The lower bound on the DL ergodic capacity in \Thmref{Theorem-Lower-Bound-Rate}, for i.i.d Rayleigh fading channels and fpZF precoding, is given by
\begin{equation}
\mathrm{SE}_k^{ \mathrm{fpZF} }\!=\!\xi^{\mathrm{DL}}\!\left(1\!-\!\frac{\tp}{\tc}\!\right)\!\log_2 \left(1\!+\!\mathrm{SINR}_k^{ \mathrm{fpZF}}\!\right) \quad \textrm{[bit/s/Hz]},
\end{equation}
where $\mathrm{SINR}_k^{ \mathrm{fpZF}}$ is given by
\begin{equation} \label{eq:SINR-ZF-alternative}
\frac{ (M-\tp) \left( \sum\limits_{l =1}^L \sqrt{ \rho_{l,k} \gamma_{l,k}} \right)^2 }{(M\!-\!\tp)\!\sum\limits_{t \!\in\!\cP_k\!\setminus\!\{k\}}\!\left(\!\sum\limits_{l =1}^{L}\!\sqrt{\rho_{l,t} \gamma_{l,k}}\!\right)^{2}\!+\!\sum\limits_{l=1}^L\!\sum\limits_{t=1}^K\!\rho_{l,t}\!\left(\!\beta_{l,k}\!-\!\gamma_{l,k}\!\right)\!+\!1}.
\end{equation}
\end{corollary}
\begin{IEEEproof}
The proof is omitted due to space limitations, but follows the same principles as in \cite[Appendix C]{Chien2016a}.
\end{IEEEproof}

\subsection{Max-Min Fairness Power Control}
The power transmitted by AP $l$ is given by
\begin{equation} \label{eq:Transmit-Power}
\EX{\norm{\bx_l}^2} = \sum_{k=1}^{K} \rho_{l,k} \EX{\norm{\bw_{l,\ik}}^2} = \sum_{k=1}^K \rho_{l,k}.
\end{equation}
Let $P_{\mathrm{max},l}$ be the maximum power that can be utilized at each AP, the per-AP power constraint is $\sum_{k=1}^{K} \rho_{l,k} \leq P_{\mathrm{max},l}$, $\forall l$.
Max-min fairness power control provides uniform SE throughout the network by allocating the DL power such that the smallest SE is maximized, at the expense of the UEs with good channel conditions. This policy gives higher relative priority to smaller SEs. Mathematically speaking, the max-min optimization problem can be formulated as
\begin{equation} \label{Problem:Max-Min-QoS-Original}
\begin{aligned}
& \underset{ \{ \rho_{l,k} \geq 0 \}}{\textrm{maximize}} & & \min_{k} \mathrm{SE}_k^{\mathrm{fpZF}} \\
& \textrm{subject to} & & \sum_{k=1}^{K} \rho_{l,k} \leq P_{\mathrm{max},l}, \; \forall l,\\
\end{aligned}
\end{equation}
which in turn is equivalent to maximize the lowest SINR value, i.e., $\displaystyle\min_{k}~\mathrm{SINR}_k^{ \mathrm{fpZF}}$, with $\mathrm{SINR}_k^{\mathrm{fpZF}}$ given by \Eqref{eq:SINR-ZF-alternative}.

In order to write \eqref{Problem:Max-Min-QoS-Original} on epigraph form we introduce the following notation: $\bz_t = \left[\sqrt{z_{1,t}}, \ldots, \sqrt{z_{L,t}}\right]\trans $ and $\bg_t = \left[\sqrt{g_{1,t}}, \ldots, \sqrt{g_{L,t}}\right]\trans $, where $g_{i,k} =  (M-\tp) \gamma_{i,k}$ and $z_{i,k} = \beta_{i,k} - \gamma_{i,k}$. Let $\bU = [\bu_1, \ldots , \bu_K ] \in \mC^{M \times K}$ have columns $\bu_{t} = \left[\sqrt{\rho_{1,t}}, \ldots, \sqrt{\rho_{L,t}}\right]\trans, \mbox{ for } t= 1, \ldots, K$, and $i$th row denoted by $\bu_{i}'$. 
Let $\bs_k \in \mC^{K+ |\mathcal{P}_k|}$ be \[ \left[\!\sqrt{\nu} \left(\!\bg_k\trans \bu_{t_{1}^{'}},\!\ldots,\!\bg_k\trans\!\bu_{t_{\left|\!\cP_k \setminus \{k\}\!\right|}^{'}},\!\norm{\bz_k\!\circ \!\bu_1},\!\ldots,\!\norm{\bz_k\!\circ\!\bu_K},\!1\!\right)\!\right]\trans, \] where $t_{1}^{'}, \ldots, t_{|\cP_k\! \setminus \{k\} |}^{'}$ are all the UE indices belonging to $\cP_k \setminus \{k\}$, and $|\mathcal{P}_k|$ is the cardinality of the set $\mathcal{P}_k$. The operator $\circ$ indicates the Hadamard product. 

The equivalent epigraph formulation of \eqref{Problem:Max-Min-QoS-Original} is given by
\begin{equation} \label{eq:SOCP} 
 \begin{aligned}
 & \underset{ \{ \rho_{i,t} \geq 0 \}, \nu}{\textrm{maximize}}
 & & \nu \\
 & \mbox{subject to}
 & & || \mathbf{s}_k || \leq \mathbf{g}_k\trans \mathbf{u}_k, \; \forall k,\\
 & & & || \mathbf{u}_{i}'|| \leq \sqrt{P_{\mathrm{max},i}}, \; \forall i.
 \end{aligned}
 \end{equation}
In \eqref{eq:SOCP}, the constraint functions are second-order cones with respect to $\{\rho_{i,t}\}$, but jointly in $\{\rho_{i,t}\}$ and $\nu$. Consequently, \eqref{eq:SOCP} is a convex program if $\nu$ is fixed, and the optimal solution can be obtained by using interior-point toolbox CVX~\cite{cvx2012}. 
Moreover, since the SINR constraint is increasing function of $\nu$, the solution to~\eqref{eq:SOCP} is obtained by solving the feasibility problem~\eqref{eq:SOCP-2} through \textit{bisection method}~\cite{Boyd2004a}:

\begin{equation} \label{eq:SOCP-2} 
 \begin{aligned}
 & {\textrm{find}}
 & & \{ \pmb{\rho}_{k} \} \\
 & \mbox{subject to}
 & & || \mathbf{s}_k || \leq \mathbf{g}_k\trans \mathbf{u}_k, \; \forall k,\\
 & & & || \mathbf{u}_{i}'|| \leq \sqrt{P_{\mathrm{max},i}}, \; \forall i, \\
 & & & \rho_{i,t} \geq 0, \; \forall i, \forall t.
 \end{aligned}
\end{equation}
where $\pmb{\rho}_{k}=[\rho_{1,k},\ldots,\rho_{L,k}]\trans \in \mC^{L \times 1}$.

%
%

\section{Numerical Results}
We compare the SEs provided by fpZF scheme, MRT scheme assuming single-antenna APs (sMRT)~\cite{Ngo2017b} and MRT scheme assuming multi-antenna APs (mMRT)~\cite{Ngo2018a}.
 These precoding schemes
have equal fronthauling requirements.

The large-scale fading coefficients $\{\beta_{l,k}\}$ are modeled as in~\cite{Ngo2017b}, assuming uncorrelated shadow fading with standard deviation $\sigmash$. The maximum radiated power is 200 mW per AP and 100 mW per UE. The normalized transmit powers in DL and UL, denoted by $\rho_{\textrm{d}}$, $\rho_{\textrm{u}}$, respectively, are defined as in~\cite{Ngo2017b}. We assume $p_k = \rho_{\textrm{u}}$, $k=1,\ldots,K$.
To simulate a cell-free network, we wrap the simulation area around with eight identical neighbor areas.
The UL pilots are randomly assigned to the UEs, and we assume $\tp=10$. 
The simulation settings are reported in Table~\ref{tab:settings}.
\begin{table}[!t]
\caption{Simulation settings}
\label{tab:settings}
\centering
\renewcommand{\arraystretch}{1.3}
\begin{tabular}{l | l || l | l}
Description & Value & Description & Value \\
\hline
Simulation area & 500$\times$500 m$^2$ & $\tc$ (symbols) & 200\\
AP/UE distribution & unif. rand. & $\xi^{\mathrm{DL}}$ & 0.5\\
AP/UE antenna height & 15/1.65 m & $\sigmash $ & 8 dB \\
Carrier frequency & 2 GHz & Bandwidth & 20 MHz \\
Noise figure & 9 dB & $K$ & 20 \\
Coherence bandwidth & 200 kHz & Coherence time & 1 ms
\end{tabular}
\end{table}
\subsection{Performance Evaluation}
The per-AP power constraint is, regardless of the precoding scheme, given by $\E\{\norm{\bx_l}^2\} \leq P_{\mathrm{max},l} = \rho_{\textrm{d}}, \forall l=1,\ldots,L$. The total transmitted power is directly proportional to $L$.
For the fpZF scheme, the max-min fairness power control coefficients $\{\rho_{l,k}\}$ are the solutions of problem~\eqref{eq:SOCP}. For sMRT and mMRT scheme, the corresponding power control coefficients are  obtained as in~\cite{Ngo2017b,Ngo2018a}, respectively.

\begin{figure}[!t]
\centering
\includegraphics[width=\linewidth]{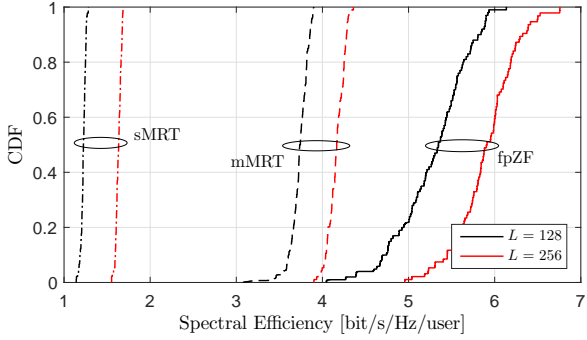} \vspace{-5mm}
\caption{CDF of the per-user SE for sMRT, mMRT, and fpZF scheme. $M=64$, $M=1$ for mMRT and sMRT scheme, respectively. $K=20$. $\tp = 10$.} \vspace{-5mm}
\label{fig:fig1}
\end{figure}

In \Figref{fig:fig1} the cumulative distribution function (CDF) of the  per-user DL SE is shown for the three precoding schemes. We consider two setups: $L=128$; $L=256$. The multi-antenna AP schemes have $M=64$, while $M=1$ for sMRT. Comparing the two MRT schemes, obviously more antennas at each AP, everything else being equal, is beneficial. fpZF outperforms mMRT of about 30\% in terms of 95\%-likely per-user SE, for both the setups. This gap derives from the inherent ability of fpZF to null the inter-user interference. \Figref{fig:fig1} also shows that fpZF can significantly outperform MRT even halving the number of APs from 256 to 128.  

\begin{figure}[!t]
\centering
\includegraphics[width=\linewidth]{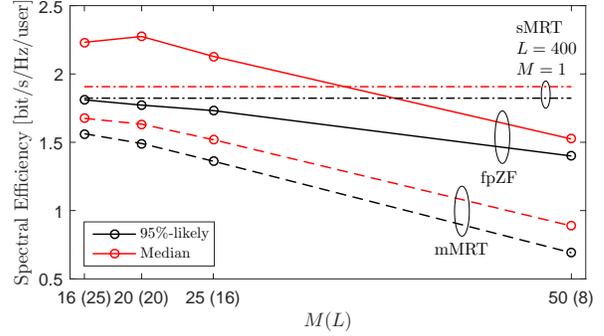} \vspace{-5mm}
\caption{Per-user SE against $M$, given $LM=400$. $K = 20$. $\tp = 10$.}
\label{fig:fig2} \vspace{-4mm}
\end{figure}

\Figref{fig:fig2} compares the SE of the multi-antenna APs schemes against $M$, for systems having the same total number of antennas, $LM = 400$, but different number of APs. For~\eqref{eq:SINR-ZF-alternative} to be meaningful, $M$ must satisfy the condition $M \geq \tp + 1$. The sMRT SE is also illustrated as reference.  
Focusing on the 95\%-likely SE, we observe that fpZF approaches the SE of sMRT as soon as $L$ is sufficiently large to guarantee a good degree of macro-diversity and $M$ is enough, compared to $\tp$, to spatially separate UEs and suppress inter-user interference. In general, increasing $L$, while keeping $LM$ fixed and satisfying $M \geq \tp + 1$, reduces the gap between the schemes (albeit fpZF stays significantly above mMRT). In addition, \Figref{fig:fig2} clearly shows the decreasing trend of the SEs as $M$ increases. These elements suggests that the macro-diversity gain is dominant over the array gain, which is $M-\tp$, $M$ for fpZF and mMRT, respectively. In terms of median SE, fpZF is, by far, the best scheme when it comes a system adequately distributed. Lastly, since the total transmitted power is proportional to the number of APs, fpZF is $L^{\mathrm{sMRT}}/L^{\mathrm{fpZF}}$ times more power-efficient than sMRT.

\begin{figure}[!t]
\centering
\includegraphics[width=\linewidth]{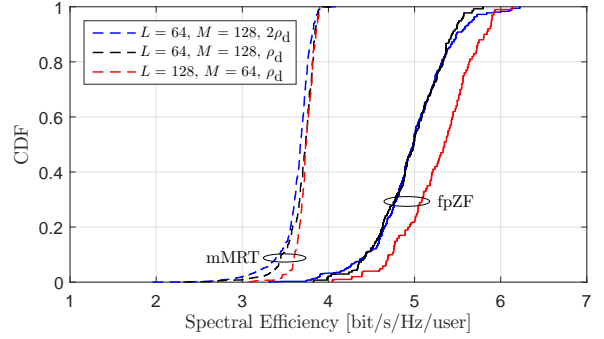} \vspace{-5mm}
\caption{CDF of the per-user SE for mMRT and fpZF, given $LM=8192$.}
\label{fig:fig3} \vspace{-5mm}
\end{figure}

\Figref{fig:fig3} confirms that the macro-diversity gain plays a more important role than the array gain. We see that fpZF gains more from the increased macro-diversity, compared to mMRT, since the channel estimation improves as the APs are closer to the UEs, and the fpZF scheme benefits more from accurate channel estimates than the mMRT scheme.
Interestingly, doubling the per-AP radiated power (blue curves) does not help at all, due to the corresponding increased interference. Hence, for a fixed $LM$, it is more convenient to distribute the total power budget over more APs rather than more antenna elements.
      

\section{Conclusion}
We analyzed the performance of fpZF scheme from \cite{Bjornson2016a} in cell-free Massive MIMO system and derived a closed-form expression for the DL SE with imperfect CSI. The simulation results show that, with max-min fairness power control, fpZF can provide higher SE than maximum-ratio transmission, without sharing CSI among the APs.


\bibliographystyle{IEEEtran}
\bibliography{IEEEabrv,refs}
\end{document}